\documentclass[amsmath,amssymb,12pt,superscriptaddress,nofootinbib]{revtex4}

\usepackage{graphicx}
\usepackage{dcolumn}
\usepackage{bm}
\begin{document}
\thispagestyle{empty}

{\baselineskip0pt
\leftline{\baselineskip14pt\sl\vbox to0pt{
               \hbox{\it Asia Pacific Center for Theoretical Physics}
               \vspace{1mm}
               \hbox{\it Osaka City  University}
               \vspace{1mm}
               \hbox{\it Yukawa Institute for Theoretical Physics} 
               \vss}}
\rightline{\baselineskip16pt\rm\vbox to20pt{
            {\hbox{APCTP-Pre2009-004}
            \hbox{OCU-PHYS-311}
            \hbox{AP-GR-65}
            \hbox{YITP-09-33}
            }
\vss}}%
}

\vskip2cm
\title{Hoop Conjecture and the Horizon Formation Cross Section
in Kaluza-Klein Spacetimes
}

\author{Chul-Moon Yoo}\email{c_m_yoo@apctp.org}
\affiliation{
Asia Pacific Center for Theoretical Physics, \\
Pohang University of Science and Technology, Pohang 790-784,~Korea
}

\author{Hideki Ishihara}\email{ishihara@sci.osaka-cu.ac.jp}
\author{Masashi Kimura}\email{mkimura@sci.osaka-cu.ac.jp}
\affiliation{ 
Department of Mathematics and Physics,
Graduate School of Science, Osaka City University,
3-3-138 Sugimoto, Sumiyoshi, Osaka 558-8585, Japan
}
\author{Sugure Tanzawa}\email{tanzawa@yukawa.kyoto-u.ac.jp}
\affiliation{
Yukawa Institute for Theoretical Physics, Kyoto University
Kyoto 606-8502, Japan
}

\date{\today}

\begin{abstract}
We analyze momentarily static initial data sets of the gravitational 
field produced by two-point sources 
in five-dimensional Kaluza-Klein spacetimes. 
These initial data sets
are characterized by the mass, the separation of sources 
and the size of a extra dimension.  
Using these initial data sets, we 
discuss the condition for black hole formation, and propose a new conjecture 
which is a hybrid of the four-dimensional hoop conjecture and 
the five-dimensional 
hyperhoop conjecture. 
By using the new conjecture, we estimate 
the cross section of black hole formation due to
collisions of particles 
in Kaluza-Klein spacetimes. 
We show that the mass dependence of the cross section 
gives us information about the size and the number of the
compactified extra dimensions. 
\end{abstract}

\pacs{04.50.+h, 04.70.Bw}

\maketitle

\pagebreak

\section{Introduction}\label{sec1}

Classical theory of gravity in higher dimensions has
gathered much attention since the brane world, 
which suggests the possibility of large extra dimensions,  has been proposed
~\cite{ArkaniHamed:1998rs,Antoniadis:1998ig}.
Black holes in this framework would be believed 
as key objects for verification of extra dimensions. 
It has been clarified that
higher-dimensional black holes in asymptotically flat spacetimes
have richer structure than four-dimensional black holes 
~\cite{Cai:2001su, Myers:1986un,Emparan:2001wn,Galloway:2005mf} 
(see also~\cite{lrr-2008-6}).
It was also  suggested that
higher-dimensional mini-black holes might be produced
in accelerators~\cite{Banks:1999gd,Dimopoulos:2001hw,Giddings:2001bu,
Ida:2002ez,PhysRevD.69.049901,Ida:2005ax,Ida:2006tf} 
and in cosmic ray events~\cite{Argyres:1998qn,Feng:2001ib,Anchordoqui:2001cg}.
Such black holes, which would evaporate by the Hawking radiation,
are expected to play crucial roles 
in the development of the quantum theory of gravity.

In this paper, we focus on 
the black hole formation rate in higher-dimensional spacetimes. 
The hoop conjecture, proposed by Thorne\cite{Thorne:1972ji}, gives 
a criterion for the black hole formation in four-dimensional spacetimes.
It is thought that the criterion by conjecture can be applied to 
a variety of the black hole formation processes. 
However, as will be mentioned in Sec.\ref{sec:hyperhoop},
the existence of black string solutions means that
Thorne's hoop conjecture, where the length 
of one-dimensional loops are used to 
measure the compactness of a system, 
is not true in higher dimensions. 
In higher-dimensional spacetimes, the hyperhoop conjecture has been proposed 
as the condition for black hole formation
~\cite{Ida:2002hg,Barrabes:2004rk,Yoo:2005nj}.
In the hyperhoop conjecture, the area of codimension three closed 
surfaces is used instead of the length of one-dimensional loops.

Apparent horizon formation is analyzed in the collision of two-point particles,
and then higher-dimensional black hole formation rates 
in accelerators has been 
predicted \cite{Yoshino:2005ps,Yoshino:2005hi,Yoshino:2002tx}.
These works are concentrated on the cases in which 
spacetimes have asymptotically Euclidean spatial sections. 
The assumption of the asymptotically Euclidean spatial sections 
is likely to be  relevant if the black hole size is much smaller than
the size of extra dimensions.
On the other hand, 
if the size of the extra dimensions are comparable to the size of 
black hole,  the formation rate would be changed. 
Does the black hole formation 
rate give us any information about 
the size of the extra dimensions?

In an asymptotically flat $D$-dimensional spacetime with 
Euclidean spatial sections, 
a typical black hole has the horizon radius $\sim(G_DM)^{1/(D-3)}$, 
where $G_D$ and $M$ are the gravitational constant 
and the mass of the black hole, respectively. 
In this case, the mass dependence of the cross section 
for black hole formation $\sigma_{\rm p}$ 
in the collision of particles is\footnote{
In brane world scenarios, 
since particles of matter are confined on a three-brane, 
it would be useful to consider the cross section which has the dimension 
of (length)$^2$.}
\begin{equation}
\sigma_{\rm p}\propto M^{2/(D-3)}, 
\end{equation}
(see Ref.~\cite{Kanti:2008eq} as a recent review, and references therein). 
If the space has compactified directions, 
this dependence should be modified. 
Suppose that $n$ directions in the $D$-dimensional spacetime 
are compactified into a length scale $l$. 
If $l\gg(G_DM)^{1/(D-3)}$, the mass dependence of $\sigma_{\rm p}$ 
is the same as the case of asymptotically flat spacetimes with  
Euclidean spatial sections. 
On the other hand, if $l\ll(G_DM)^{1/(D-3)}$, 
we expect that the compactified dimensions can be neglected, and 
$\sigma_{\rm p}$ is given by the horizon radius of 
a typical ($D-n$)-dimensional 
black hole $(G_{D-n} M)^{1/(D-n-3)}$. 
Namely, we expect that
the mass dependence of $\sigma_{\rm p}$ behaves as
\begin{equation}
	\sigma_{\rm p}\propto M^{2/(D-n-3)}. 
\end{equation}
This transition of the mass dependence of $\sigma_{\rm p}$ might give us the 
information  about the  compactification scale.

In this paper, we consider systems of two-point particles 
in a five-dimensional Kaluza-Klein spacetime. 
We use, concretely, the four-dimensional Euclidean Taub-NUT space~\cite{Gibbons:1979zt},
which has a twisted ${\rm S}^1$ as the extra dimension. 
We construct initial data sets of the gravitational field around 
two-point particles including the parameter which describes the 
separation of the particles. 
As varying the separation parameter, 
we inspect the existence of a cover-all
apparent horizon using the same technique as
is used in Ref.~\cite{Yoo:2007mq}. 
We will show that there is the maximum separation parameter 
for the existence of a cover-all apparent horizon, 
and we consider this apparent horizon indicates 
the {necessary and sufficient} compactness of the 
system for the black hole formation. 
From the shape of this apparent horizon, we will obtain the condition 
for the black hole formation in the higher-dimensional spacetime with 
the compactified extra dimension.

For the following reason, 
we consider the four-dimensional Taub-NUT space, 
which is a twisted S$^1$ fiber bundle 
over the flat three-dimensional base space, as a time slice, 
not a direct product of S$^1$ and the base space.
Let us consider a spherically symmetric black hole in a
five-dimensional asymptotically flat spacetime 
with Euclidean spatial sections. 
The geometry admits SO(4) spatial isometry. 
If we impose a periodic identification 
in a spatial direction which causes S$^1$ compactification of a direct product, 
the isometry reduces to SO(3). 
In contrast, a black hole can have SO(3)$\times$U(1) symmetry 
if it is in a five-dimensional spacetime where 
the extra dimension is the twisted S$^1$ fiber over the four-dimensional spacetime
\cite{Dobiasch:1981vh,Gibbons:1985ac,Ishihara:2005dp}. 
Similarly, in the systems of two-point sources, 
the symmetry is U(1) in the direct product spaces while it can be 
U(1)$\times$U(1) in the twisted S$^1$ bundle cases. 
The spaces with twisted S$^1$ bundle structure can have more symmetry 
than the simple direct product spaces. 
Using this advantage, 
recently, black hole solutions with nontrivial asymptotic structure 
are studied in the five-dimensional 
Einstein-Maxwell theory~\cite{Nakagawa:2008rm,Tomizawa:2008hw,Tomizawa:2008rh, Ishihara:2006iv,
Ishihara:2006pb,Kimura:2008cq,Ishihara:2006ig,Ida:2007vi,Matsuno:2007ts,Tomizawa:2008qr}. 
This advantage makes it possible to 
search for apparent horizons 
by solving ordinary differential equations 
in the space with twisted S$^1$ bundle structure\cite{Yoo:2007mq}.

In a five-dimensional Kaluza-Klein spacetime, 
we propose the new condition of horizon formation 
which is a hybrid of the four-dimensional hoop conjecture and the
five-dimensional hyperhoop conjecture. 
Extrapolating the new proposal to general situations, 
we estimate the cross section of the black hole formation 
in collision of particles as a function of the mass scale 
in any dimension.
We show that 
the mass dependence of the cross section 
changes when the mass scale becomes comparable to the scale of 
the extra dimension.

The organization of the paper is as follows. 
The method for constructing the initial data sets 
is shown in Sec.\ref{sec2}. 
In Sec.\ref{sec3}, the hyperhoop conjecture  
in the spacetime with a compactified extra dimension is 
examined, and a new conjecture is proposed. 
Effects of the compactification size of the extra dimensions
on the black hole production cross section 
are discussed in Sec.\ref{sec5}, and 
summary and discussions are given in Sec.\ref{sec6}.

\section{Momentarily static initial data in Kaluza-Klein spaces}
\label{sec2}
In this section, 
as a preparation to discuss the hyperhoop conjecture in Kaluza-Klein 
spaces,
we construct initial data sets
for two-point sources 
with a compactified extra dimension, 
and discuss geometrical properties. 

\subsection{Construction of initial data}
Let us consider an initial data set of 
the induced metric and the extrinsic curvature $(h_{ij},K_{ij})$ 
on a four-dimensional spacelike hypersurface $\Sigma$, 
which satisfies 
the Hamiltonian and momentum constraints,
\begin{align}
	{\cal R}-K_{ij}K^{ij}+K^2&=16\pi G_5\rho_{\rm m},\\
	D_j\left(K^{ij}-h^{ij}K\right)&=8\pi G_5J_{\rm m}^i,
\end{align}
where $\rho_{\rm m}$ and $J_{\rm m}^i$ 
are the energy density and the energy flux of matter, 
and $D_i$ and ${\cal R}$ are the 
covariant derivative within $\Sigma$ and the scalar curvature with respect to 
$h_{ij}$.

We restrict ourselves to momentarily static cases, i.e.,
\begin{eqnarray}
	K_{ij} &=& 0, 
\end{eqnarray}
and assume the induced metric has the form of
\begin{eqnarray}
	h_{ij} dx^i dx^j &=& F^2 ds_{\rm RF}^2,
\label{eq:confRF}
\end{eqnarray}
where $ds_{\rm RF}^2$ is a Ricci flat metric. 
In this case, the vacuum momentum constraint is trivially 
satisfied and 
the vacuum Hamiltonian constraint reduces to 
\begin{equation}
	\triangle_{\rm RF}F=0,  \label{eq:raprf} 
\end{equation}
where $\triangle_{\rm RF}$ is the Laplace operator of the Ricci flat metric.

For the purpose of considering two-point sources in a Kaluza-Klein 
space, we take the two-center Taub-NUT metric, which is Ricci flat.
The metric in the Gibbons-Hawking(GH) form\cite{Gibbons:1979zt} is given by
\begin{eqnarray}
	ds_{\rm GH}^2 &=& V^{-1}
	ds_{\rm 3dE}^2
		+\frac{V}{4}l^2
		\left(d\psi+\omega_\phi d\phi\right)^2,
 \label{eq:GH} 
\\ 
ds_{\rm 3dE}^2 &=& dr^2+r^2d\theta^2+r^2\sin^2\theta d\phi^2, 
\label{eq:3deuclid}
\\
	V^{-1}&=&1+\frac{l}{2}\left(\frac{1}{\sqrt{r^2+a^2-2ar\cos\theta}}
	+\frac{1}{\sqrt{r^2+a^2+2ar\cos\theta}}\right), 
\\
	\omega_\phi&=&\frac{r\cos\theta-a}{\sqrt{r^2+a^2-2ar\cos\theta}}
	+\frac{r\cos\theta+a}{\sqrt{r^2+a^2+2ar\cos\theta}}. 
\label{eq:omega}
\end{eqnarray}
The range of angular coordinates are 
$0\leq\theta\leq\pi$, $0\leq\phi\leq2\pi$ and $0\leq\psi\leq 4\pi$.
We consider two-point sources locate at the two centers of \eqref{eq:GH} 
which are fixed points of the action of isometry 
generated by the Killing vector $\partial_\psi$. 
Then, we can assume that the function $F$ has the same symmetry, 
i.e., $F$ does not depend on $\psi$. 
In this case, the Eq.(\ref{eq:raprf}) reduces to 
\begin{equation}
	\triangle_{\rm 3dE}F=0, \label{eq:3e}
\end{equation}
where $\triangle_{\rm 3dE}$ is the Laplace operator 
on the three-dimensional Euclidean metric of \eqref{eq:3deuclid}. 
A solution of \eqref{eq:3e} for two-point sources is
\begin{equation}
	F=1+\frac{m_1/l}{\sqrt{r^2+a^2-2ar\cos\theta}}
	+\frac{m_2/l}{\sqrt{r^2+a^2+2ar\cos\theta}}. 
\label{eq:harmonics}
\end{equation}

In the limit $r\rightarrow \infty$, we have
$ F \to 1,~V^{-1} \to 1, \omega_\phi \to 2\cos\theta$, then  
we can see the asymptotic form of $h_{ij}$ as 
\begin{equation}
	h_{ij}dx^idx^j \to dr^2+r^2(d\theta^2+\sin^2\theta d\phi^2)
	+l^2 \left(\frac{d\psi}{2}+\cos\theta d\phi\right)^2. 
\label{eq:metfar}
\end{equation}
Thus, we can regard the extra dimension, twisted S$^1$ spanned by $\psi$, 
is compactified in the size $l$ at the asymptotic region. 
An $r=$constant surface of the space 
with the metric (\ref{eq:metfar}) is 
homeomorphic to the
lens space $L(2;1)={\rm S}^3/{\mathbb Z}_2$ 
\cite{Ishihara:2006ig,Ishihara:2006pb,Yoo:2007mq}.

The Abbott-Deser mass\cite{Abbott:1981ff} of the initial metric
(\ref{eq:confRF}) with 
(\ref{eq:GH})-(\ref{eq:omega}) and (\ref{eq:harmonics})
can be calculated as 
\begin{equation}
	G_5M =3\pi(m_1+m_2), 
\label{eq:admass2}
\end{equation}
where we have used the metric
\begin{align}
ds^2=\left(1+\frac{l}{r}\right) \left(dr^2
+r^2d\theta^2+
		r^2\sin^2\theta d\phi^2\right)
+ l^2 \left(1+\frac{l}{r}\right)^{-1}\left(\frac{d\psi}{2}+\cos\theta d\phi\right)^2,
\label{eq:spacemetgps}
\end{align}
which is the $a=0$ case of \eqref{eq:GH}, 
as the reference metric
\footnote{
The reference metric is different from the one used 
in Refs.\cite{Kurita:2007hu,Kurita:2008mj,Cai:2006td} 
because the topology of $r={\rm const.}$ surface at infinity is 
not ${\rm S^3}$ but ${\rm S^3}/{\mathbb Z_2}$ in our case.
The reference metric
is the same as the induced metric of a static slice 
in the Gross-Perry-Sorkin(GPS) monopole solution
\cite{Gross:1983hb,Sorkin:1983ns}
except for the factor $1/2$ in front of 
$d\psi$. 
}.
For simplicity, hereafter, we set 
\begin{equation}
	m_1=m_2=m. 
\end{equation}
The mass parameter $m$ has the dimension of length square. 
Hereafter, a nondimensional parameter $m/l^2$ is the key parameter. 

\subsection{Apparent horizon}
\label{sec:ah}

If a spacetime is an asymptotically predictable spacetime from
a Cauchy surface, and the null energy condition is satisfied,
then the existence of an apparent horizon guarantees the existence
of an event horizon \cite{Hawking:1973uf}.
Then, the existence of an apparent horizon is a 
relevant indicator for the formation of a black hole.

Before considering the $a\neq0$ cases, 
it is useful to see the $a=0$ case, 
where we can calculate the horizon radius analytically.
Putting $a=0$ in (\ref{eq:confRF}) with 
(\ref{eq:GH})-(\ref{eq:omega}) and (\ref{eq:harmonics}), we have 
the induced metric in the form 
\begin{eqnarray}
	h_{ij}dx^idx^j &=&
	\left(1+\frac{2m}{lr}\right)^2
	\bigg[
	\left(1+\frac{l}{r}\right)
	\left(dr^2 +r^2d\theta^2+ r^2\sin^2\theta d\phi^2\right)
	\notag\\ & & 
	+\left(1+\frac{l}{r}\right)^{-1}l^2
	\left(\frac{d\psi}{2}+\cos{\theta} d\phi\right)^2
	\bigg]. 
\end{eqnarray}
Then, due to the symmetry, the horizon is 
given as an $r={\rm const.}$ surface. 
The area of an $r={\rm const.}$ three-dimensional surface $A(r)$ is given by
\begin{equation}
A(r)=8\pi^2F^3V^{-1/2}r^2l=\frac{8\pi^2\sqrt{l+r}(2m+lr)^3}{l^2r^{3/2}}. 
\end{equation}
Since the initial hypersurface is momentarily static, i.e., $K_{ij}=0$, 
the apparent horizon is a
minimal surface. 
Then, 
we can obtain the horizon radius $r_{\rm h}$ as a solution of the equation
\begin{equation}
	\left.\frac{dA(r)}{dr}\right|_{r=r_{\rm h}}=0. 
\end{equation}
The horizon radius $r_{\rm h}$ 
is given by
\begin{equation}
	r_{\rm h}=\frac{1}{8l}\left[-3 l^2+4m+\sqrt{9 l^4+72m l^2+16m^2}\right]
\label{eq:rh}
\end{equation}
and 
evaluated as 
\begin{equation}
	r_{\rm h}\simeq \left\{
\begin{array}{cc}
\dfrac{m}{l}
~~&{\rm for}~~\dfrac{m}{l^2} \gg 1,
\\
\dfrac{2m}{l}
~~&{\rm for}~~\dfrac{m}{l^2} \ll 1.
\end{array}\right.
\end{equation}

In the $a\neq 0$ cases, the geometry given by 
(\ref{eq:confRF}) with 
(\ref{eq:GH})-(\ref{eq:omega}) and (\ref{eq:harmonics})
becomes less symmetric compared to the $a= 0$ case. 
But, it still has U(1)$\times$U(1) isometry generated by the commuting 
Killing vectors $\partial_\phi$ and $\partial_\psi$. 
Then an apparent horizon, if there exists, should be given by the surface 
in the form
\begin{equation}
	r = r_{\rm h}(\theta).
\end{equation}
This simplicity comes from the fact that we use the two-center Taub-NUT space 
as the base space. 
So, we have to solve the ordinary differential equation 
for minimal surfaces in the form
\begin{eqnarray}
    r_{\rm h}''-\frac{3{r_{\rm h}'}^2}{r_{\rm h}}
     -2r_{\rm h} 
      + (r_{\rm h}^2 +{r_{\rm h}'}^2) 
		\left\{ \frac{r_{\rm h}'}{r_{\rm h}^2} \cot\theta
 	- \left(
   G_r(r_{\rm h},\theta)  - \frac{r_{\rm h}'}{r_{\rm h}^2} G_\theta(r_{\rm h},\theta)
            \right)\right\} =0,
   \label{eq:ode}
\end{eqnarray}
where a prime means the derivative 
with respect to $\theta$ and functions $G_r(r,\theta)$ and $G_\theta(r,\theta)$ are
\begin{eqnarray}
G_r(r,\theta) &:= \partial_r 
          \left(3\ln F+   \frac{1}{2}\ln V^{-1} \right),
\\
G_\theta(r,\theta) &:= \partial_\theta 
          \left(3\ln F+   \frac{1}{2}\ln V^{-1} \right).
\end{eqnarray}

\begin{figure}[htbp]
\begin{center}
\includegraphics[scale=0.3]{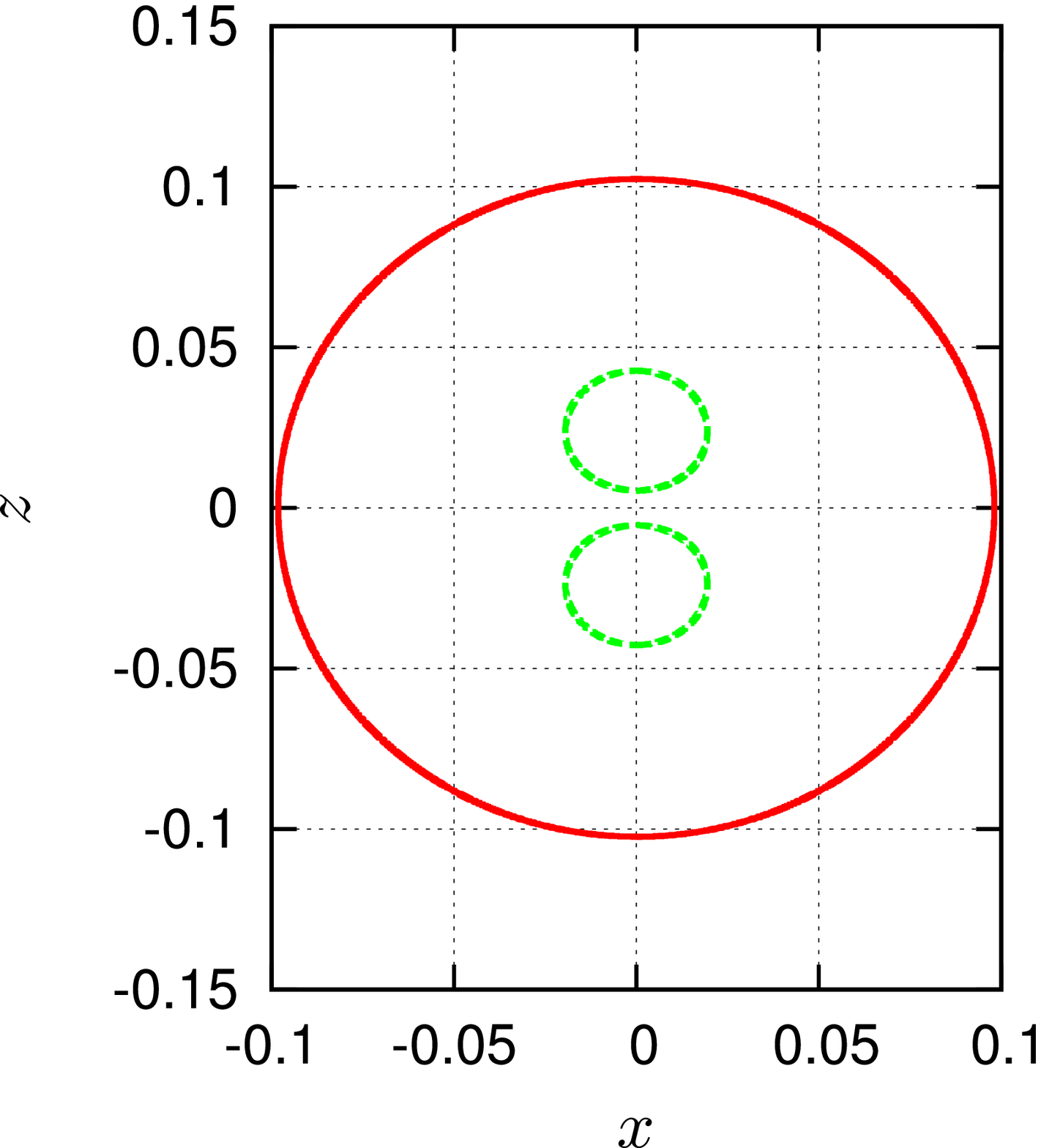}
\includegraphics[scale=0.3]{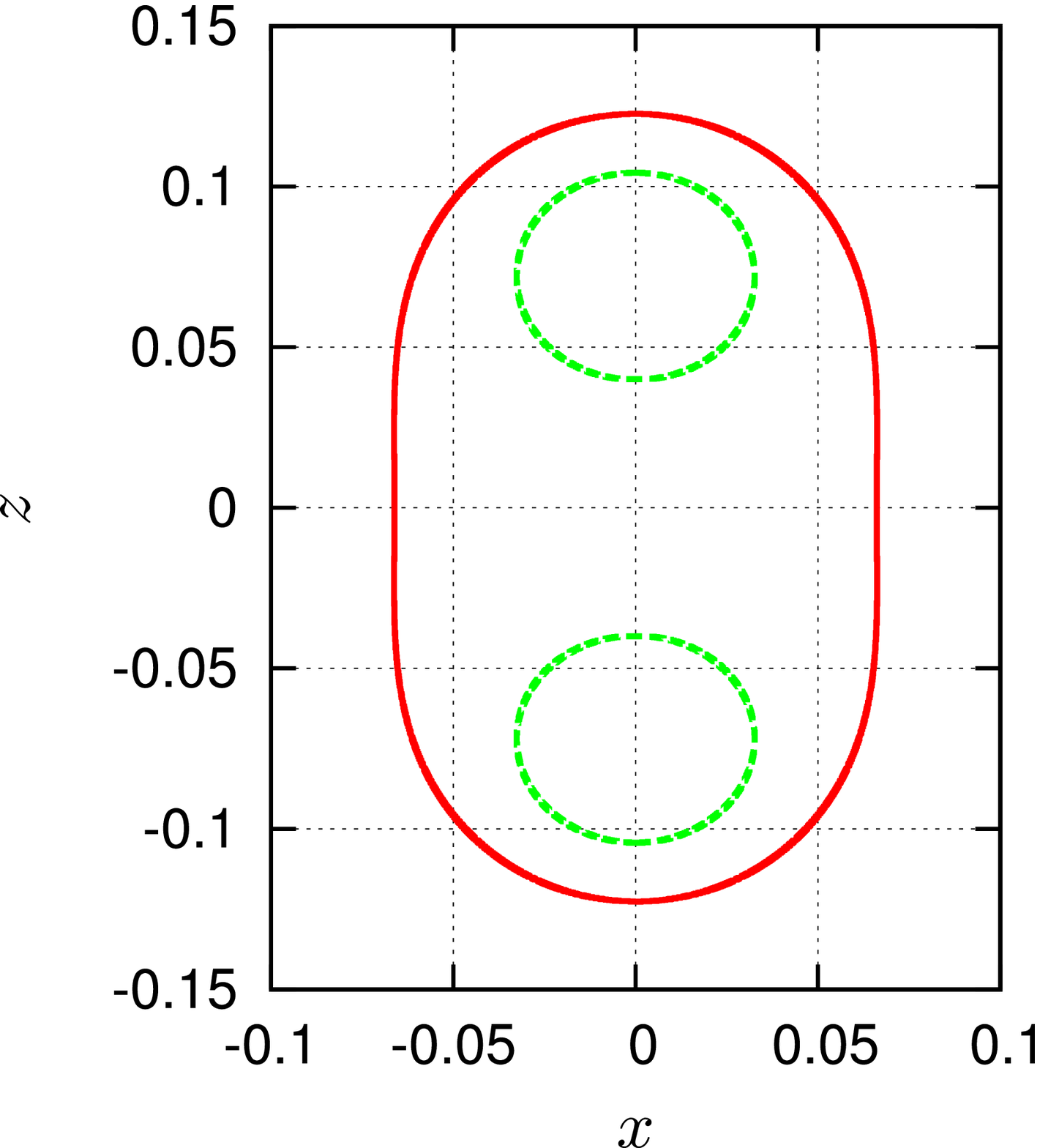}
\includegraphics[scale=0.3]{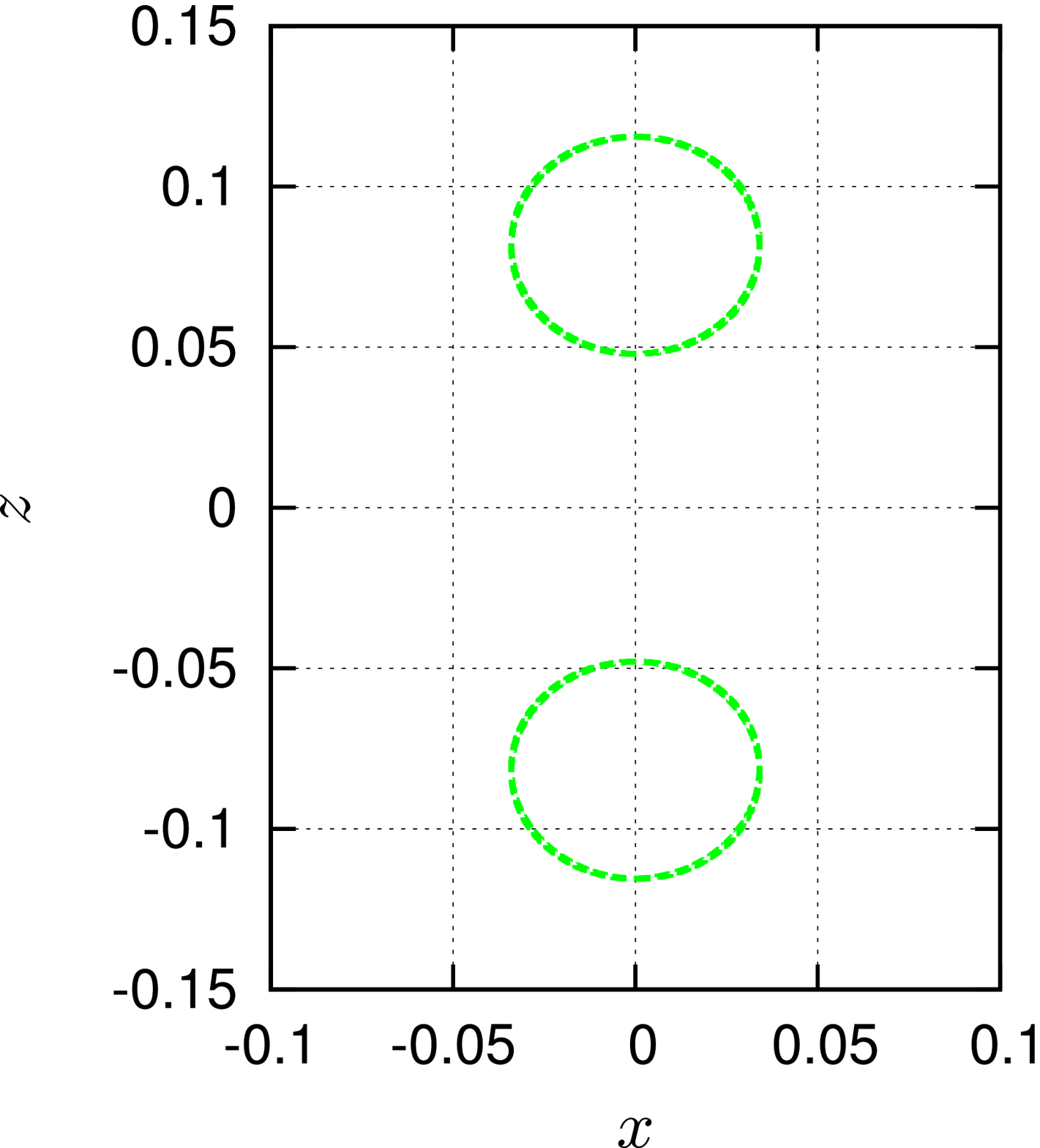}
\caption{
Apparent horizons for $m^2/l=0.0025$ and $a=0.02m^{1/2}$, $0.07m^{1/2}$ 
and $0.08m^{1/2}$ from left to right. 
Solid lines represent cover-all horizons, and dashed lines 
represent each independent horizons of two-point sources. 
In the right panel, where $a=0.08m^{1/2}$, 
no cover-all horizon exists. 
The horizontal and vertical axes represent 
$x=r\sin{\theta}$ and $z=r\cos{\theta}$, respectively.
}
\label{fig:ah}
\end{center}
\end{figure}

The apparent horizons, which are solutions of \eqref{eq:ode}, 
are described by 
smooth closed curves in two-dimensional plane $(r, \theta)$.
Typical graphs of the apparent horizons 
are plotted in Fig.\ref{fig:ah}.  
As shown in Fig.\ref{fig:ah}, 
the separation of the two-point sources must be
smaller than a certain value for the existence of 
a cover-all apparent horizon.

\section{Test of hyperhoop conjecture in Kaluza-Klein spaces}
\label{sec3}

\subsection{Hyperhoop conjecture}
\label{sec:hyperhoop}

The black hole production rate due to 
collisions of particles 
can be evaluated by using the notion of the hyperhoop conjecture for
asymptotically Minkowski spacetimes in higher dimensions\cite{Ida:2002hg,Barrabes:2004rk,Yoo:2005nj}. 
Here we check whether this conjecture is true in
the case of higher dimensions with compactified directions.

The hyperhoop conjecture 
is as follows:
{\it 
Black holes with horizons form when 
and only when a mass $M$ gets compacted into a region whose 
$(D-3)$-dimensional volume
in every direction is
\begin{equation}
	V_{D-3}\lesssim \alpha_D G_DM,
\label{eq:hhc}
\end{equation}
where $\alpha_D$ is a numerical factor and 
$G_D$ is the gravitational constant 
in $D$-dimensional theory of gravity, and the $(D-3)$-dimensional 
volume $V_{D-3}$ means the volume of a
$(D-3)$-dimensional closed submanifold(hyperhoop) of a spacelike
hypersurface.
}

It should be noted that this conjecture has 
some ambiguities.
The definitions of the mass and the hyperhoop are not 
explicitly given.
In this paper, we interpret $M$ as the total mass of a system, 
and $V_{D-3}$ as a typical $(D-3)$-dimensional volume of 
a closed submanifold 
which represents the compactness of a system.

In Thorne's original hoop conjecture, 
the one-dimensional circumference is used as an indicator of the
compactness of a system. 
However, in the five-dimensional Einstein gravity, 
we know that the black string solutions have hoops with 
infinite length. 
In addition, D. Ida and K. Nakao showed that 
the one-dimensional circumference of the apparent horizon which is
produced by a uniform line source can be infinitely long. 
Then, they proposed the hyperhoop which measures the compactness of the 
system \cite{Ida:2002hg}.

If the extra dimensions have finite sizes, 
it is nontrivial whether the volume of 
the hyperhoop can give us 
the appropriate criterion for black hole formation or not. 
When the size of a black hole is much smaller than the size of 
extra dimensions, the hyperhoop conjecture 
would be true. 
However, when the black hole is as large as the extra dimensions, 
validity of the hyperhoop conjecture is not clear. 
Then, we check whether the hyperhoop works in Kaluza-Klein spaces 
in the next subsection. 

\subsection{Test of hyperhoop conjecture in Kaluza-Klein spaces}

The authors in Ref.\cite{Barrabes:2004rk,Yoo:2005nj} studied the criterion 
of black hole formation in relation 
to the compactness of explicit matter source distributions. 
In contrast, we consider 
systems consist of only two-point masses, 
in which geometrical information is only the distance between them. 
The (hyper) hoop conjecture claims that 
if a black hole exists, any length 
scale characterizing the black hole should be 
less than a critical scale determined 
by the mass of the black hole. 
To check this statement,
we use the size of a
cover-all apparent horizon, 
if it exists, to measure the typical length scale 
of a black hole, 
and use the Abbott-Deser mass as the total mass $M$.

We have constructed the initial data of the gravitational field 
of two-point sources with the separation parameter $a$ 
in the previous section.  
Now, we discuss the criterion of black hole formation 
by introducing a geometrical quantity $V(a)$, 
which measures the compactness of a cover-all apparent horizon. 
At present, we do not restrict the dimension of $V(a)$. 

We require the inequality
\begin{align}
	V(a) \lesssim (\mbox{ critical size})
\label{eq:hhcan}
\end{align}
for cover-all apparent horizons if they exist, where the
\lq\lq critical size\rq\rq\ 
is a quantity related to the Abbott-Deser mass. 

According to the numerical calculations, there exists 
a critical value of separation parameter $a_{\rm cr}$ such that 
cover-all horizon exists if $a<a_{\rm cr}$. 
Then, we can expect that 
$V(a)$ and the ``critical size" satisfy following two properties:
\begin{itemize}
	\item[(i)]{$V(a)$ is a monotonic increasing function of $a$ at least in 
		the vicinity of $a_{\rm cr}$.}
	\item[(ii)]{$V(a_{\rm cr})\sim $(critical size). }
\end{itemize}

First, we check the hyperhoop conjecture in the form \eqref{eq:hhc} 
for $D=5$. The quantity $V(a)$ in the left-hand side of \eqref{eq:hhcan} 
is the area of a two-dimensional closed surface $V_2(a)$. 
We consider closed geodetic 2-surfaces 
$\cal A$ on a cover-all horizon which characterize the shape of horizon. 
We take the surface which has maximum area among $\cal A$ 
as $V_2(a)$.

We fix the 
\lq\lq critical size\rq\rq\ in \eqref{eq:hhcan}, 
using 
the horizon radius of five-dimensional Schwarzschild black holes 
\begin{align}
	r_{\rm Sch}=\sqrt{\frac{8G_5M}{3\pi}}. 
\label{eq:rsch}
\end{align}
Setting \eqref{eq:hhcan} holds equality 
in the case of Schwarzschild black holes, i.e., 
\begin{align}
	\mbox{critical size}= 4\pi r_{\rm Sch}^2 
=\frac{32}{3}G_5M, 
\end{align}
we have 
\begin{align}
	&V_2(a)\lesssim \frac{32}{3}G_5M. 
\label{eq:hhc2}
\end{align}
Then the numerical value of $\alpha$ in \eqref{eq:hhc} 
is also fixed as 32/3.

Because the  
geometries have the isometry group generated by 
$\partial_\phi$ and $\partial_\psi$, and 
the discrete isometry $\theta \to \pi-\theta$, 
we consider the following
typical closed geodetic 2-surfaces $\cal A$ 
on a horizon:
\begin{eqnarray}
{\cal A}_{\theta =\pi/2}
&:&{\rm area~of~}{\theta =\frac{\pi}{2}~}{\rm surface},
\\
{\cal A}_{\phi =0}
&:&{\rm area~of~}{\phi =0~}{\rm surface},
\\
{\cal A}_{\psi =0}
&:&{\rm area~of~}{\psi =0~}{\rm surface}.
\end{eqnarray}

As is noted before, we define 
\begin{eqnarray}
	V_2(a)&=&
		\max\big\{
			{\cal A}_{\theta =\pi/2},~
			{\cal A}_{\phi =0},~
			{\cal A}_{\psi =0}
		\big\}.
\label{defv2gh}
\end{eqnarray}

\begin{figure}[htbp]
\begin{center}
\includegraphics[width=0.95\linewidth,clip]{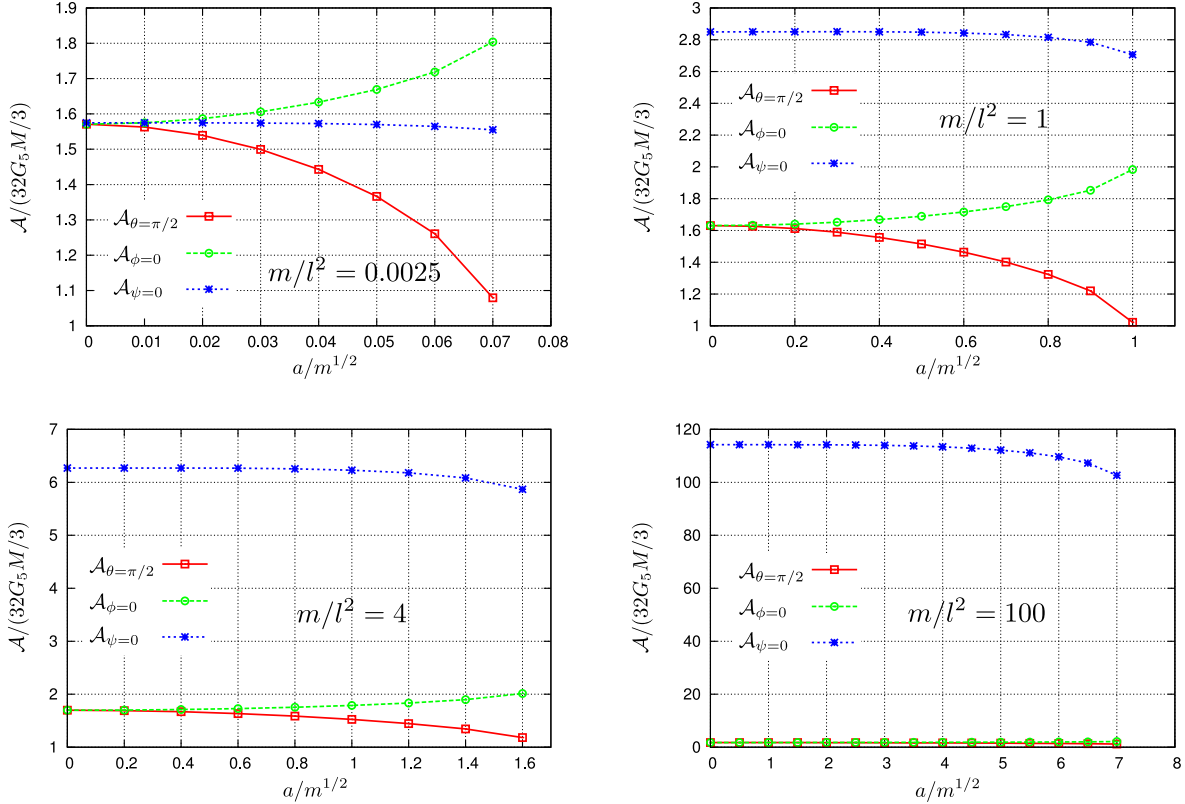}
\caption{ 
Areas of the two-dimensional geodetic surfaces 
$\mathcal A$ on apparent horizons are depicted as 
functions of $a/m^{1/2}$ in the cases of 
$m/l^2=0.0025$, 1, 4, and 100. 
$ V_2=\mathcal A_{\psi=0}$, the maximum value of $\mathcal A$ is 
not a monotonic increasing function of $a/m^{1/2}$ 
in the cases $m/l^2= 1,4,100$. 
The critical values of ${\cal A}_{\psi =0}$ become 
much larger than the critical size $32 G_5M/3$ in 
these cases. 
}
\label{fig:V2gh}
\end{center}
\end{figure}

The values of $\cal A$'s 
are depicted as functions of $a$ in some cases  
of $m/l^2$ in Fig.\ref{fig:V2gh}. 
%
%
%
As can be seen in the Fig.\ref{fig:V2gh}, 
$V_2={\cal A}_{\phi =0}$ in the 
$m/l^2 = 0.0025$
case and 
$V_2={\cal A}_{\psi =0}$ in the other cases.
In the cases of $m/l^2 \gtrsim 1$, 
$V_2$ is not monotonic increasing function of $a$. 
We can see also $V_2(a_{\rm cr})$ becomes much 
larger than the critical size, $32 G_5M/3$. 
Namely, both of conditions (i) and (ii) are not satisfied. 
In contrast, 
we show that these two conditions are satisfied in the 
asymptotically Euclidean case in Appendix\ref{sec:af}. 

This failure of the hyperhoop conjecture may be clear 
if we consider a direct product
spacetime $S^1 \times M_{\rm Sch}$, where $M_{\rm Sch}$ is a four-dimensional
Schwarzschild spacetime.
In this spacetime, when $G_5M/l^2>1$, $V_2$ is given by
\begin{align}
V_2\sim 16\pi(G_4M)^2\sim 4G_5^2M^2/(\pi l^2), 
\end{align}
where we have used the relation between $G_4$ and $G_5$ given by 
\begin{align}
	G_4\sim \frac{G_5}{2\pi l}. 
\label{eq:G}
\end{align}
Though $V_2/(G_5M)$ can be infinitely large for $G_5M/l^2\gg1$, 
a horizon exists for any $G_5M/l^2$. 
This fact means the failure of the condition (ii).

It should be noted that we cannot give 
the complete counterexample for the hyperhoop conjecture 
in a nontrivial asymptotic structure in this paper. 
Because there are ambiguities in 
the definition of $V_{D-3}$ and the mass which should be used in 
the Eq.(\ref{eq:hhc}), and also in the 
interpretation of the ``$\lesssim$". 
Nevertheless, we found the completely different
feature from the asymptotically Euclidean case which suggests 
the hyperhoop conjecture cannot be 
extended straightforwardly to cases with finite sizes of extra dimensions.

\subsection{Criterion for large black hole formation}

Next, 
let us focus on the $m/l^2 \gg 1$ 
case, where $V_2$ is not appropriate for the 
left-hand side of the condition (\ref{eq:hhc}). 
In this case, 
the size of the black hole is much larger than that of the extra dimension, and
the gravitational field  outside the horizon is effectively four-dimensional. 
Then, we can expect the ordinary four-dimensional hoop conjecture 
\begin{align}
	V_1\lesssim 4\pi G_4 M 
\label{eq:hc}
\end{align}
is true, where the constant $\alpha$ has been determined by using four-dimensional 
Schwarzschild black holes. 
Then we take one-dimensional hoop $V_1(a)$ as $V(a)$ in \eqref{eq:hhcan}. 
Using Eq.(\ref{eq:G}), 
we rewrite \eqref{eq:hhcan} as
\begin{align}
	V_1(a) \lesssim \frac{2G_5 M}{l}.\label{eq:hcv1}
\end{align}

In order to estimate $V_1(a)$, we consider 
the following
typical closed geodesic curves $\cal C$ 
on  a horizon:  
\begin{eqnarray}
&{\cal C}^{\phi =0}_{\psi =0} 
&:{\rm length~of~}{\phi =0}{\rm ~and~}{\psi =0~}{\rm curve},
\\
&{\cal C}^{\theta =\pi/2}_{\psi =0}
&:{\rm length~of~}{\theta =\frac{\pi}{2}}{\rm ~and~}{\psi =0~}{\rm curve},
\\
&{\cal C}^{\theta =0}_{\psi =0}
&:{\rm length~of~}{\theta =0}{\rm ~and~}{\psi =0~}{\rm curve},
\\
&{\cal C}^{\theta =\pi/2}_{\phi =0}
&:{\rm length~of~}{\theta =\frac{\pi}{2}}{\rm ~and~}{\phi =0~}{\rm curve},
\end{eqnarray}
and we define 
\begin{eqnarray}
	V_1(a)&=&
	\max\big\{
	{\cal C}^{\phi =0}_{\psi =0},~
	{\cal C}^{\theta =\pi/2}_{\psi =0},~
	{\cal C}^{\theta =0}_{\psi =0},~
	{\cal C}^{\theta =\pi/2}_{\phi =0}
	\big\}. 
\label{eq:defv1gh}
\end{eqnarray}
Here, we have taken the isometry of the horizon geometry into account 
as before. 
If $V_1$ gives  a measure for black hole formation, the properties (i) and (ii) 
should be satisfied.

\begin{figure}[htbp]
\begin{center}
\includegraphics[width=0.95\linewidth,clip]{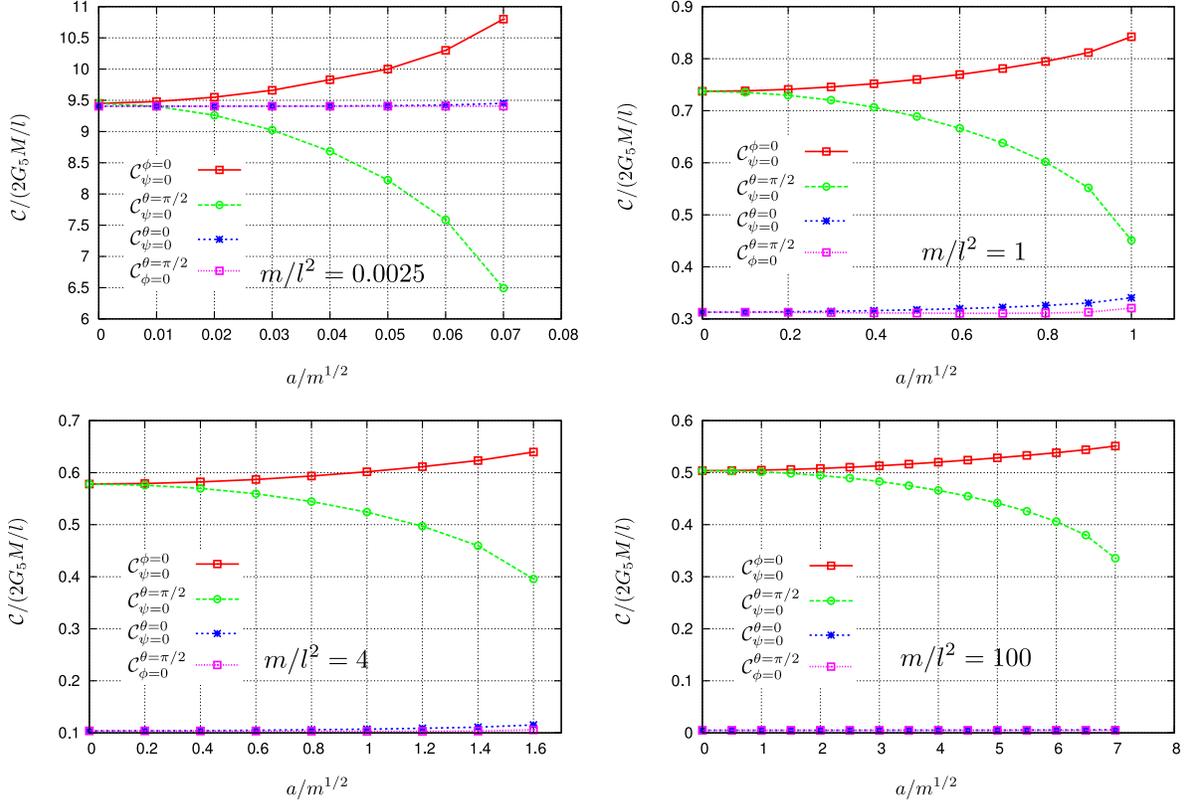}
\caption{Length of closed one-dimensional geodesic curves 
$\mathcal C$ on apparent horizons 
are depicted as functions of $a/m^{1/2}$ in 
the cases of $m/l^2=0.0025$, 1, 4, and 100. 
The critical values of $V_1=\mathcal C^{\phi=0}_{\psi=0}$ 
become much larger than the critical size $2G_5M/l^2$ in the 
$m/l^2=0.0025$ case. 
}
\label{fig:V1gh}
\end{center}
\end{figure}

The values of $\cal C$'s 
are depicted as functions of $a$ in some cases  
of $m/l^2$ in Fig.\ref{fig:V1gh}.
As can be seen in the Fig.\ref{fig:V1gh}, 
the maximal one is ${\cal C}^{\phi =0}_{\psi =0}$ in all cases, 
then $V_1={\cal C}^{\phi =0}_{\psi =0}$. 
Though $V_1$ is  a monotonic increasing function of $a$ for all 
cases of $m/l^2$, 
$V_1(a_{\rm cr})$ becomes 
much larger than the critical size, $2G_5 M/l$, 
in the 
case $m/l^2 \ll 1$.

\begin{table}[htbp]
\caption{Inequalities and properties 
for black hole formation. 
}
\label{tab:summ}
\begin{tabular}{|c||c|c|c|c||c|c|}
\hline
Inequality&
\multicolumn{2}{c|}{$ V_2\lesssim 32 G_5M/3 $}&
\multicolumn{2}{c||}{$ V_1\lesssim 2G_5 M / l$}
&\multicolumn{2}{c|}{$W\lesssim 32G_5M/3$}\\
\hline
\hline
Property&(i)&(ii)&(i)&(ii)&(i)&(ii)\\
\hline
$m/l^2 \gg 1 $&
\makebox[2cm][c]{No}&\makebox[2cm][c]{No}&
\makebox[2cm][c]{Yes}&\makebox[2cm][c]{Yes}&
\makebox[2cm][c]{Yes}&\makebox[2cm][c]{Yes}\\
\hline
$m/l^2 \ll 1 $&
\makebox[2cm][c]{Yes}&\makebox[2cm][c]{Yes}&
\makebox[2cm][c]{Yes}&\makebox[2cm][c]{No}&
\makebox[2cm][c]{Yes}&\makebox[2cm][c]{Yes}\\
\hline
\end{tabular}
\end{table}

The results of the test are summarized in Table \ref{tab:summ}. 
We find from this table that 
$V_2$ gives  an appropriate measure for the criterion of horizon formation 
only for $m/l^2 \lesssim 1$, 
while  $V_1$ does only for $m/l^2 \gtrsim 1$. 
%

\subsection{Hybrid condition}

Since $V_2$ works in the cases $m/l^2 \lesssim 1$, 
and $V_1$ works in the cases $m/l^2 \gtrsim 1$, 
we can expect that a combination of $V_1$ and $V_2$ 
provides a good measure in all range of the mass scale 
for horizon formation in Kaluza-Klein spaces. 
According to the results in the previous subsection, 
we can immediately find 
$V_2\gg lV_1$ for $m/l^2\gg 1$ and 
$V_2\ll lV_1$ for $m/l^2\ll 1$. 
Then we propose
a new condition for horizon formation: 
\begin{equation}
	W\lesssim 
	\frac{32}{3}G_5M
\label{eq:whhc}
\end{equation}
with the following definition of $W$:
\begin{align}
	\frac{1}{W}:=\frac{3}{16lV_1}+\frac{1}{V_2}. 
\label{eq:defw}
\end{align}

\begin{figure}[htbp]
\begin{center}
\includegraphics[scale=1]{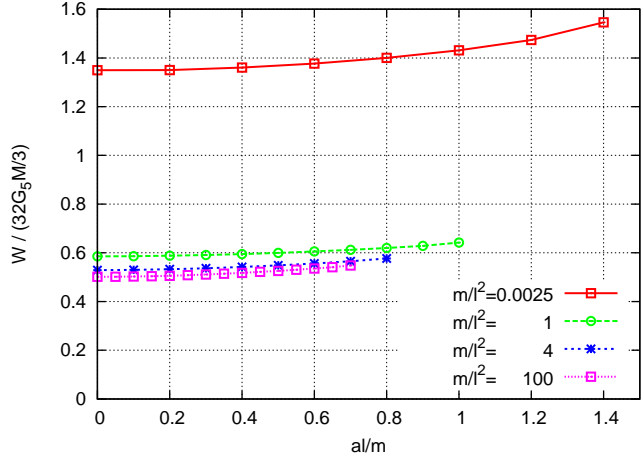}
\caption{Values of $W$ are depicted as functions of $al/m$ for the cases 
of $m/l^2=0.0025$, 1, 4, 100. 
$W$ is a monotonic increasing function of $al/m$ and 
$W(a_{\rm cr}) \sim 1$ in all cases. 
}
\label{fig:tv2}
\end{center}
\end{figure}

We plot $W$ as a function of $a$ in Fig.\ref{fig:tv2}. 
We can see from this figure that $W$ satisfies two properties: 
(i) $W$ is a monotonic increasing function of $a$; 
and (ii) $W(a_{\rm cr}) \sim \frac{32}{3}G_5M$. 
Of course, in the asymptotically Euclidean case, i.e., $l \to \infty$, 
\eqref{eq:whhc} reduces to the inequality \eqref{eq:hhc2}, 
and in the limit $l \to 0$ it reduces to \eqref{eq:hcv1}. 
Then, the condition \eqref{eq:whhc} with \eqref{eq:defw} is a hybrid 
of the four-dimensional hoop conjecture and the five-dimensional 
hyperhoop conjecture\footnote{
In fig.\ref{fig:tv2}, $W$ does not tend to 1 in $a \to 0$ limit even though 
$m \ll l^2$ or $m \gg l^2$. 
This would be because the horizon topology is not S$^3$. Furthermore, 
the geometry of the initial surfaces
differs from a time slice of squashed black hole solutions 
in the case $m \gg l^2$. }. 

We should note that there is a significant difference between $W$ and $V_2$. 
The hyperhoop $V_2$ is just a geometrical quantity 
which represents the typical size of the horizon, 
while $W$ contains the size of extra dimension $l$ 
which is related to the asymptotic property.

Extrapolating this idea to general dimensions $D$, 
we can consider the following extended version of the
hyperhoop conjecture:
{\it Black holes with horizons form in a
$D$-dimensional spacetime
when 
and only when a mass M gets compacted into a region whose 
$n$-dimensional volume $V_n$ ($n=1,2,...,D-3$) 
in every direction satisfy
\begin{align}
	\left(\sum_i^n\frac{1}{\beta_iV_i\prod^{D-i-3}_kl_k}
	\right)^{-1}\lesssim G_DM,
	\label{eq:anydimhybrid}
\end{align}
where $\beta_i$ are numerical factors and 
$l_n, ~(l_1\leq l_2 \cdots \leq l_{D-4})$ are the compactification scales 
of each compactified direction, 
and the $n$-dimensional volume means the volume of a $n$-dimensional 
closed submanifold of a spacelike hypersurface. 
}

\section{cross section of black hole production in Kaluza-Klein spaces}
\label{sec5}

In this section, 
on the basis of the new conjecture proposed in the previous section,
we discuss the mass dependence of the cross section of 
black hole formation 
due to collision of particles. 

\subsection{Case of five dimensions}
\begin{figure}[htbp]
\begin{center}
\includegraphics[scale=1.2]{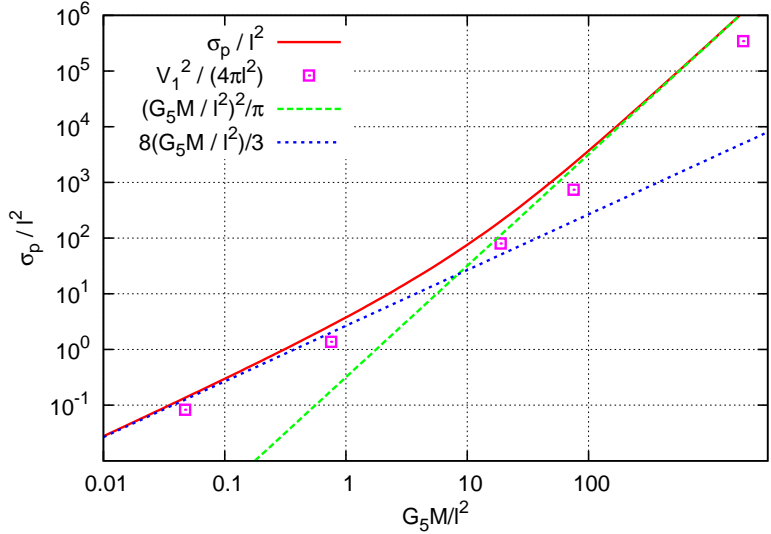}
\caption{
The cross section $\sigma_{\rm p}$ of five-dimensional Kaluza-Klein 
black hole formation which is estimated by the 
hybrid hoop conjecture is depicted as a function of
$G_5M/l^2$. 
}
\label{fig:sigma}
\end{center}
\end{figure}

In the five-dimensional case, 
the black hole 
formation condition could be given by 
\eqref{eq:whhc} with \eqref{eq:defw}. 
In the case of the collision of particles, we expect that 
the shape of a black hole is not highly elongated in our 
four dimensions. 
Then, the cross section $\sigma_{\rm p}$ likely to be given by 
\begin{align}
	\sigma_{\rm p}\sim \pi\left(\frac{V_1}{2\pi}\right)^2
					\sim \pi\left(\frac{V_2}{4\pi}\right). 
	\label{eq:sigma_p}
\end{align}
Based on this assumption, we can estimate $\sigma_{\rm p}$ 
using \eqref{eq:whhc} with \eqref{eq:defw} as follows. 
We replace $V_1$ and $V_2$ in \eqref{eq:defw} by using \eqref{eq:sigma_p}, 
and set $W=\frac{32}{3}G_5M$, then we get
\begin{align}
	\frac{3}{16l\sqrt{4\pi\sigma_{\rm p}}}+\frac{1}{4\sigma_{\rm p}}
	=\frac{3}{32G_5M}. 
\end{align}
We can solve this equation with respect to $\sigma_{\rm p}$ as 
\begin{align}
	\sigma_{\rm p}/l^2
	=\frac{8}{3}G_5M/l^2 +\frac{1}{2\pi}(G_5M/l^2)^2
		+ \frac{1}{2\pi}G_5M/l^2
	\sqrt{\frac{32\pi}{3} G_5M/l^2+(G_5M/l^2)^2}. 
\label{eq:defsigma}
\end{align}
The value of $\sigma_{\rm p}$ 
is plotted in Fig.\ref{fig:sigma} as a function of 
$G_5M/l^2$. 
The numerical values of $V_1^2(a_{cr})/4\pi$ 
for five different values
of $m/l^2=G_5M/(6\pi l^2)$ are superposed 
in Fig.\ref{fig:sigma}. 

We can see a transition of power-law dependence of 
the cross section on the mass scale from $\sigma_{\rm p}\propto G_5M$  
to $\sigma_{\rm p}\propto (G_5M)^2$ 
as $M$ increases. 
The total mass can be regarded as the center of mass energy 
in the high energy particle collision. 
The mass dependence of the cross section comes from the mass dependence 
of size and shape of the horizons.\footnote{Actually, 
we can see the mass dependence of apparent horizon size of the 
initial data in the $a=0$ case as shown in Appendix \ref{ap4}. }
Although the values of $V_1^2(a_{cr})/4\pi$ are somewhat deviated 
from the line of $\sigma_{\rm p}$, 
these are still same order, and 
the plots follow the transition of $\sigma_{\rm p}$.

\subsection{Case of arbitrary dimension}

It is interesting to consider a compactification 
which has different compactification scales.
At least, one of these scales contributes
to resolving the hierarchy problem,
then it should be much larger than the length scale
$\sim 10^{-17}$cm for TeV gravity scenarios.
We can consider a possibility that other compactification scales take intermediate scales.
It would be possible that the energy scale of colliding particles 
may be the same order of the intermediate compactification scales. 
In this case, 
we can obtain the information of the numbers of compactified dimensions and 
the size of them from the mass dependence of the black hole production rate.

To see this, let us consider that compact $n_*$ dimensions have a size $l_*$, 
compact $n_{\rm L}$ dimensions, 
which would contribute to resolving the hierarchy problem, 
are larger than $l_*$, 
and other compact dimensions are much smaller than $l_*$. 
For a black hole with mass $M$, if its horizon size is smaller than $l_*$, 
the effective dimension is $D_{\rm eff}=4+n_{\rm L}+n_*$, 
i.e., we should consider 
the black hole is in a $(4+n_{\rm L}+n_*)$-dimensional spacetime, 
where we assumed the sizes of other extra dimensions are much 
smaller than the horizon size. 
On the other hand, if the horizon is larger than $l_*$, 
the effective dimension 
is $D_{\rm eff}=4+n_{\rm L}$ because $n_*$ dimensions become ineffective, 
i.e., the black hole is effectively in $(4+n_{\rm L})$ dimensions.   

The condition of black hole formation in $D_{\rm eff}$ dimensions would be 
given by
\begin{equation}
	V_{D_{\rm eff}-3}\lesssim \alpha_{D_{\rm eff}}G_{D_{\rm eff}} M,
\end{equation}
where
\begin{equation}
	\alpha_{D}=\frac{16\pi}{D-2}\frac{{\Omega}_{D-3}}{{\Omega}_{D-2}}, 
\end{equation}
and ${\Omega}_D$ is the $D$-dimensional area of the unit $D$-sphere. 
For smaller black holes, $D_{\rm eff}=4+n_{\rm L}+n_*$, and 
for larger black holes $D_{\rm eff}=4+n_{\rm L}$. 
As is done in \eqref{eq:sigma_p}, we estimate $\sigma_{\rm p}$ by 
\begin{eqnarray}
	&&V_{(1+n_{\rm L}+n_*)} 
		= \left(\frac{\sigma_{\rm p}}{4\pi}\right)^{(1+n_{\rm L}+n_*)/2} 
			{\Omega}_{(1+n_{\rm L}+n_*)}, \cr
	&&V_{(1+n_{\rm L})}
		= \left(\frac{\sigma_{\rm p}}{4\pi}\right)^{(1+n_{\rm L})/2} 
			{\Omega}_{(1+n_{\rm L})}. 
			\label{eq:vsigmap}
\end{eqnarray}
Then, the mass dependence of the cross section for small black holes 
is 
\begin{equation}
	\sigma_{\rm p}
		\simeq 4\pi\left(\frac{16\pi G_{(4+n_{\rm L}+n_*)}}{(2+n_{\rm L}+n_*){\Omega}_{(2+n_{\rm L}+n_*)}} 
			M\right)^{2/(1+n_{\rm L}+n_*)}, 
\label{eq:SmallBH}
\end{equation}
and for large black holes 
\begin{equation}
	\sigma_{\rm p}
		\simeq4\pi\left(\frac{16\pi G_{(4+n_{\rm L})}}{(2+n_{\rm L}){\Omega}_{(2+n_{\rm L})}} 
			M\right)^{2/(1+n_{\rm L})}. 
\label{eq:LargeBH}
\end{equation}
Therefore, by $\log \sigma_{\rm p}-\log M$ plot, 
we obtain directly the numbers of dimensions $(1+n_{\rm L}+n_*), (1+n_{\rm L})$, and 
effective gravitational constants  $G_{(4+n_{\rm L}+n_*)}$ and $G_{(4+n_{\rm L})}$. 
Then, we can estimate 
$l_*$ and the volume of  the extra dimensions larger than $l_*$, 
say  ${\rm Vol}_{n_{\rm L}}$, by 
\begin{eqnarray}
	&&2\pi l_* \sim \left(\frac{G_{(4+n_{\rm L}+n_*)}}{G_{(4+n_{\rm L})}}\right)^{1/n_*}, \cr
	&&{\rm Vol}_{n_{\rm L}} \sim \frac{G_{(4+n_{\rm L})}}{G_4} .
\end{eqnarray}
>From the crossover point $M_*, \sigma^*_{\rm p}$ of the cross section 
from \eqref{eq:SmallBH} to \eqref{eq:LargeBH} 
we can also estimate $l_*$ and ${\rm Vol}_{n_{\rm L}}$ as
\begin{eqnarray}
	2\pi l_{\rm *}\hspace{1mm}
	&\sim&  \left(\frac{2+n_{\rm L}+n_*}{2+n_{\rm L}}
\frac{\Omega_{(2+n_{\rm L}+n_*)}}{\Omega_{(2+n_{\rm L})}}\right)^{1/n_*}
		\left(\frac{\sigma_{\rm p}^*}{\pi}\right)^{1/2},
		\label{eq:lstar}
		\\
	{\rm Vol}_{n_{\rm L}} 
		&\sim& \frac{(2+n_{\rm L})\Omega_{(2+n_{\rm L})}}{16\pi G_4 M_*}
		\left(\frac{\sigma_{\rm p}^*}{\pi}\right)^{(1+n_{\rm L})/2}
		. 
\end{eqnarray}
As a demonstration, in 
the case of total dimension $D=10$ and compact dimension 6, we show the $M$ dependence of $\sigma_{\rm p}$ which 
is defined by using Eqs.(\ref{eq:vsigmap}) 
and the saturated inequality (\ref{eq:anydimhybrid})
(see Fig.6). 
\begin{figure}[htbp]
\begin{center}
\includegraphics[scale=0.9]{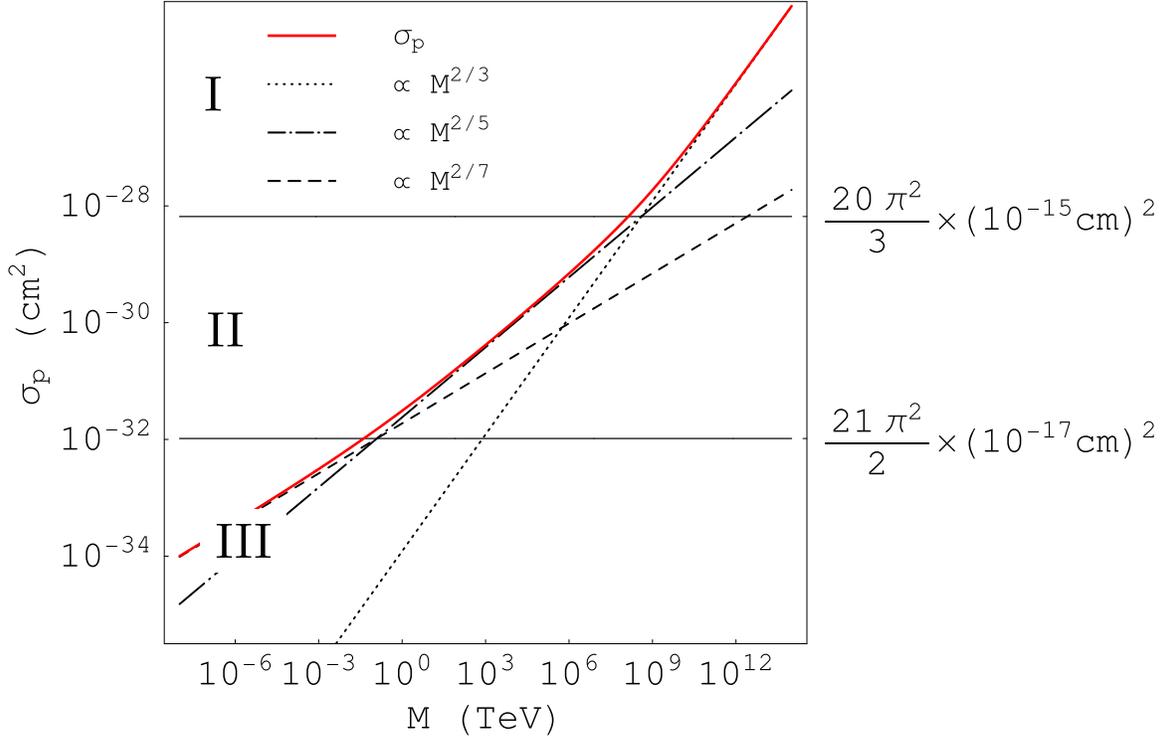}
\caption{
The cross section of 
black hole formation $\sigma_{\rm p}$, which is estimated by the 
hybrid hoop conjecture, is depicted as a function of 
the center of mass energy $M$. 
The total dimension is 10 and six of them are compactified in 
each scale. 
Two directions, which 
contribute to resolving the hierarchy problem, 
are compactified in $10^{-2}$cm. 
Another two of the compactified directions have the size 
$10^{-15}$cm, and the remaining two dimensions have the size 
$10^{-17}$cm. 
The power exponent of the cross section 
changes around $\sigma_{\rm p}=20\pi^2/3\times (10^{-15}{\rm cm})^2$ 
and $\sigma_{\rm p}=21\pi^2/3\times (10^{-17}{\rm cm})^2$ which 
 are given by Eq.(\ref{eq:lstar}). 
The numbers of effective dimensions $D_{\rm eff}$ are given by 
6, 8, 10 in the regions I, II, and III, respectively. 
The power exponent of $\sigma_{\rm p}$ in 
each region is 
given by ${2}/{(D_{\rm eff}-3)}$ on the mass scale. 
}
\label{fig:2stage}
\end{center}
\end{figure}

\section{Summary and Discussions}
\label{sec6}

We have constructed initial data sets of the gravitational field 
produced by two-point masses 
which represent Kaluza-Klein spaces in the asymptotically far region. 
These systems are characterized by the size $l$ of the compactified dimension,
the mass scale $m$, and  the separation of the particles $a$. 
By using these initial data we have investigated the geometry of 
apparent horizons and 
condition of horizon formation in the five-dimensional 
Kaluza-Klein spacetimes. 
Furthermore, we have discussed 
the cross section of black hole production 
by two particle systems.

Thorne's original hoop conjecture represents that 
an one-dimensional closed curve $V_1$ gives the condition of 
black hole formation in four-dimensional spacetimes. 
The hyperhoop conjecture represents that 
a two-dimensional closed surface $V_2$ gives the condition in 
five-dimensional spacetimes. 
These would be true 
if the spacetimes have asymptotically Euclidean spatial sections.
However, we have shown that the hyperhoop conjecture in five dimensions 
is not valid in the case of an asymptotically Kaluza-Klein space where 
the size of an extra dimension $l$ is comparable to the scale 
of the black hole horizon.
Instead, we have proposed an alternative condition 
[see \eqref{eq:whhc}] 
for horizon formation 
using a geometrical value $W$ which is defined as the harmonic average of 
$l V_1$ and $V_2$. 
The exact definition is given by Eq.\eqref{eq:defw}.
Then the new criterion
is a hybrid of the four-dimensional hoop conjecture and 
the five-dimensional hyperhoop conjecture. 
The quantity $W$ contains not only the quantities 
$V_1$ and $V_2$ but also 
the size $l$ of the compactified dimension, 
which is related to the asymptotic geometry.

Using the value of $W$, 
we have investigated the mass dependence of 
the cross section $\sigma_{\rm p}$ of black hole 
formation [see \eqref{eq:defsigma}]. 
As expected, in five dimensions, 
$\sigma_{\rm p}$ is proportional to the mass 
when the mass scale is much less than the scale of 
the compact dimension, 
$G_5M/l^2 \ll 1$,  
while $\sigma_{\rm p}$ is in proportion to the mass square 
when the mass scale is much larger than the scale of the
compact dimension, $G_5M/l^2 \gg 1$. 
We have shown that the transition of $\sigma_{\rm p}$ actually occurs. 
We can obtain the information of the size of extra dimensions 
by observing the mass dependence of the black hole production rate. 
If the total dimension is larger than 5, and 
extra dimensions would have different compactification scales, 
we expect from the hybrid hoop conjecture that 
the mass dependence of the black hole production rate tells us 
the number of large compact dimensions and 
the each size of them.

Because the larger energy makes the larger size of 
a black hole, 
then it gives us the information of larger extra dimensions. 
Even though the energy scale in laboratories is too small to 
verify this effect, we hope that the information of the large extra dimensions 
is given by active phenomena in astrophysics through the mass 
dependence of the black hole production rate.

\section*{Acknowledgement}
We are grateful to Dr.Yasunari Kurita, Professor Ken-ich Nakao, 
and Professor Misao Sasaki 
for helpful discussions and comments. 
This work is supported in part by JSPS 
Grant-in-Aid for Scientific Research
No. 19540305 and for Scientific Research (B) No. 17340075.
MK is supported by the JSPS Grant-in-Aid for Scientific Research No.
20$\cdot$7858.

\appendix
\section{Test of hyperhoop conjecture in asymptotically Euclidean spaces}
\label{sec:af}
Let us consider the case of asymptotically Euclidean spaces. 
We can write the conformally flat induced metric as 
\begin{align}
	h_{ij}dx^idx^j
	=F^2\left[dr^2+r^2\left(d\theta^2+\sin^2\theta \left(d\phi^2
	+\sin^2\phi ~d\psi^2\right)\right)\right], 
\end{align}
where the range of the angular coordinates is given by 
$0\leq\theta\leq\pi$, $0\leq\phi\leq\pi$, and $0\leq\psi\leq 2\pi$.

The vacuum Hamiltonian constraint becomes 
\begin{equation}
	\triangle_{\rm 4dE}F=0, \label{eq:4e}
\end{equation}
where $\triangle_{\rm 4dE}$ is the 
Laplace operator on the four-dimensional 
Euclidean metric. 
A solution of this equation which has two-point sources is given by
\begin{equation}
	F=1+\frac{2m_1}{r^2+a^2-2ar\cos\theta}+\frac{2m_2}{r^2+a^2+2ar\cos\theta}, 
\end{equation}
where $m_1, m_2$ are mass parameters of each particle
and $a$ is the separation parameter. 
For this initial data, we can calculate 
the ADM mass 
as 
\begin{equation}
	G_5M_{AD}=3\pi (m_1+m_2). 
\end{equation}

\begin{figure}[htbp]
\begin{center}
\includegraphics[width=0.5\linewidth,clip]{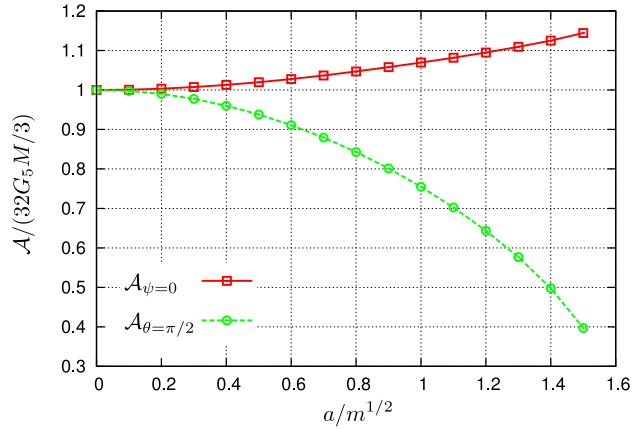}
\caption{$\mathcal A$: Areas of the two-dimensional geodetic surfaces on 
apparent horizons. 
}
\label{fig:V2eu}
\end{center}
\end{figure}

Hereafter, we set $m_1=m_2=m$. 
Using the initial data, we obtain 
apparent horizons $r=r_{\rm h}(\theta)$ by 
the same way in Sec.\ref{sec2}. 
We consider the following typical closed geodetic 2-surfaces on a 
horizon: 
\begin{eqnarray}
	{\cal A}_{\theta =\pi/2}
		&:&{\rm area~of~}{\theta =\frac{\pi}{2}~}{\rm surface},
\\
	{\cal A}_{\psi =0}
		&:&{\rm area~of~}{\psi =0~}{\rm surface}, 
\end{eqnarray}
and the hyperhoop $V_2(a)$,  
\begin{eqnarray}
	V_2(a)&=&
		\max\big\{
		{\cal A}_{\theta =\pi/2},~
		{\cal A}_{\psi =0}
		\big\}.
\label{v2def2}
\end{eqnarray}
The values of ${\cal A}$'s are depicted as functions of $a/\sqrt{m}$ in 
Fig.\ref{fig:V2eu}. 

%

As shown in this figure,  
$V_2= {\cal A}_{\psi =0}$ 
is a monotonic increasing function of $a$. 
In addition,  
$3 V_2(a_{\rm cr})/
(32 \pi G_5 M)
\sim 1$. 
That is, $V_2(a)$ satisfies the two properties (i) and (ii) in the text.

\section{Mass dependence of the horizon size 
 in the $a=0$ case}
\label{ap4}
In the $a=0$ case, as noted in Sec.\ref{sec2}, we can easily find
the apparent horizon. 
We can calculate the areas of the closed geodetic 2-surfaces $\cal A$'s and 
the length of the closed geodesics $\cal C$'s on the apparent horizon 
analytically, as Table \ref{tab:geocb}. 
In Table \ref{tab:geocb},
$r_{\rm h}$ is given by \eqref{eq:rh} and 
we have carefully chosen the integral ranges of the angular coordinates 
for the surfaces and the curves to be closed. 
\begin{table}[htbp]
\caption{Closed geodetic surfaces, where
$
F_{a=0}=1+\frac{2m}{lr_{\rm h}}$, 
$V_{a=0}^{-1}=1+\frac{l}{r_{\rm h}} $, 
and
$E[x]$ is the complete elliptic integral of second kind
defined by 
$
E[x]:=\int^{\frac{\pi}{2}}_0\sqrt{1-x\sin^2\theta}d\theta. 
$
}
\label{tab:geocb}
\begin{tabular}{|c||c|c|c|c|}
\hline
Name&\makebox[3cm][c]{$\mathcal C^{\phi=0}_{\psi=0}$}
&\makebox[3cm][c]{$\mathcal C^{\theta=\pi/2}_{\psi=0}$}
&\makebox[3cm][c]{$\mathcal C^{\theta=0}_{\psi=0}$}
&\makebox[3cm][c]{$\mathcal C^{\theta=\pi/2}_{\phi=0}$}\\
\hline
Definition&$\psi$, $\phi=$0
&$\theta=\pi/2$, $\psi=$0
&$\theta=0$, $\psi=$0 
&$\theta=\pi/2$, $\phi=$0\\
\hline
Period&$0\leq\theta\leq2\pi$
&$0\leq\phi\leq2\pi$
&$0\leq\phi\leq2\pi$
&$0\leq\psi\leq4\pi$\\
\hline
\hline
Length ($a=0$)&$2\pi r_{\rm h}F_{a=0} {V_{a=0}}^{-1/2}$
&$2\pi r_{\rm h}F_{a=0} {V_{a=0}}^{-1/2}$
&$2\pi lF_{a=0} {V_{a=0}}^{1/2}$
&$2\pi lF_{a=0} {V_{a=0}}^{1/2}$\\
\hline
\hline
$m/l^2\gg 1$&$6\pi m/l$  
&$6\pi m/l$  
&$6\pi l$  
&$6\pi l$  \\
\hline
$m/l^2\ll 1$&$4\pi\sqrt{2m}$  
&$4\pi\sqrt{2m}$  
&$4\pi\sqrt{2m}$  
&$4\pi\sqrt{2m}$  \\
\hline
\end{tabular}
\medskip

\begin{tabular}{|c||c|c|c|}
\hline
Name&\makebox[4cm][c]{$\mathcal A_{\theta=\pi/2}$}
&\makebox[4cm][c]{$\mathcal A_{\phi=0}$}
&\makebox[4cm][c]{$\mathcal A_{\psi=0}$}\\
\hline
Definition&$\theta=\pi/2$
&$\phi=$0
&$\psi=$0\\
\hline
Period&$0\leq\phi\leq2\pi$, $0\leq\psi\leq4\pi$
&$0\leq\theta\leq2\pi$, $0\leq\psi\leq4\pi$
&$0\leq\theta\leq2\pi$, $0\leq\phi\leq2\pi$\\
\hline
\hline
Area ($a=0$)&$4\pi^2r_{\rm h}l{F_{a=0}}^2$
&$4\pi^2r_{\rm h}l{F_{a=0}}^2$
&$8\pi r_{\rm h}l{F_{a=0}}^2E\left[-\frac{r_{\rm h}}{l}\left(\frac{r_{\rm h}}{l}+2\right)
\right]$
\\
\hline
\hline
$m/l^2\gg 1 $&$36\pi^2m$  
&$36\pi^2m$  
&$72\pi (m/l)^2$  \\
\hline
$m/l^2\ll 1$&$32\pi^2m$  
&$32\pi^2m$  
&$32\pi^2m$ \\
\hline
\end{tabular}
\end{table}

The 
horizon is a squashed lens space, and 
the ratio of the size of S$^1$ fiber to the size of S$^2$ base is 
given by 
\begin{eqnarray}
\frac{\mathcal C^{\theta=0}_{\psi=0}}{\mathcal C^{\phi=0}_{\psi=0}}
=\left(1+\frac{l}{r_{\rm h}}\right)^{-1}\frac{l}{r_{\rm h}}. 
\end{eqnarray}

If we consider the case of $m/l^2 \gg 1$, 
then $r_{\rm h} \simeq m/l$. 
The hoop $V_1$ and the hyperhoop $V_2$ given by (\ref{eq:defv1gh}) 
and (\ref{defv2gh}) behave as
\begin{eqnarray}
	V_1 & \simeq & \frac{6\pi m}{l}, 
\\
	V_2 & \simeq & \frac{72\pi^2 m^2}{l^2}. 
\end{eqnarray}

On the other hand, if we consider the case of 
$m/l^2 \ll 1$, then $r_{\rm h} \simeq 2m/l$ and 
\begin{align}
V_1 & \simeq {\cal C}^{\phi =0}_{\psi =0}\Big|_{a=0}
={\cal C}^{\theta =\pi/2}_{\psi =0}\Big|_{a=0}
\simeq{\cal C}^{\theta =0}_{\psi =0}\Big|_{a=0}
={\cal C}^{\theta =\pi/2}_{\phi =0}\Big|_{a=0}
\simeq  4\pi \sqrt{2 m},\\
V_2 & \simeq 
{\cal A}_{\theta =\pi/2}\Big|_{a=0}
={\cal A}_{\phi =0}\Big|_{a=0}
\simeq {\cal A}_{\psi =0}\Big|_{a=0}
\simeq  32 \pi^2 m. 
\end{align}
They mean that the apparent horizon is round.

\end{document}